# Instrumental and scientific simulations of the LOFT Wide Field Monitor

Y. Evangelista[*a,b], I. Donnarumma[a], R. Campana[c,d], C. Schmid[e], M. Feroci[a,b]
on behalf of the LOFT collaboration

[a]INAF-IAPS, Via del Fosso del Cavaliere 100, I-00133 Roma, Italy; [b]INFN Roma Tor Vergata, Via della Ricerca Scientifica 1, I-00133, Roma, Italy; [c]INAF-IASF Bologna, via Gobetti 101, I- 40129 Bologna, Italy; [d]INFN Bologna, Viale Berti Pichat 6/2, I-40127 Bologna, Italy; [e]Dr. Karl-Remeis-Sternwarte and Erlangen Centre for Astroparticle Physics (ECAP), Sternwartstr. 7, 96049, Bamberg, Germany

## ABSTRACT

The Large Observatory for X-ray Timing (LOFT) is one of the five candidates that were considered by ESA as an M3 mission (with launch in 2022-2024). It is specifically designed to exploit the diagnostics of very rapid X-ray flux and spectral variability that directly probe the motion of matter down to distances very close to black holes and neutron stars, as well as the physical state of ultradense matter. The LOFT scientific payload is composed of the Large Area Detector (LAD), devoted to spectral-timing observation, and the Wide Field Monitor (WFM), whose primary goal it is to monitor the X-ray sky for transient events that need to be followed up with the LAD, and to measure the long-term variability of galactic X-ray sources and localize gamma-ray bursts. Here we describe the simulations carried out to optimize the WFM design and to characterize the instrument response to both isolated sources and crowded fields in the proximity of the galactic bulge.

## 1. INTRODUCTION

The LOFT[1] WFM is a coded aperture imaging experiment designed on the heritage of the SuperAGILE[2] experiment, successfully operating in orbit since 2007.[3]

With the ~120 μm position resolution provided by its single-sided Silicon microstrip detector, SuperAGILE demonstrated the feasibility of a compact, large-area, light, low-power and high resolution X-ray imager, with steradian-wide field of view. The LOFT WFM applies the same concept, with improvements provided by the superior performance of Silicon Drift Detectors (SDDs) used in place of the Si microstrips. These detectors provide a lower energy threshold and better energy resolution, and represent the state-of-the art in large area, monolithic Silicon detectors. In a coded mask instrument, the mask shadow projected by the observed sources is recorded by the position-sensitive detector and can be deconvolved by using proper reconstruction procedures to recover the image of the sky.[4]

In this paper we describe the results obtained by means of a mixed analytic/Monte Carlo simulator specifically developed to optimize the instrument design and to evaluate the scientific performance of the LOFT Wide Field Monitor.

In the following we will describe the LOFT/WFM design (Section 2), the software used for the simulation of astrophysical point sources and cosmic X-ray background (Section 3.1), the simulation of the SDD based detection plane (Section 3.2), the image reconstruction algorithm (Section 3.3) and the imaging results (Section 3.4 and 3.5). Finally, in Section 4, we draw our conclusions.

---

[*] yuri.evangelista@iaps.inaf.it, (+39) 06 4993 4657

## 2. LOFT/WFM DESIGN

The LOFT WFM camera has physical dimension of ~30 × 30 × 30 cm$^3$, with a mass of ~7 kg and a power dissipation of 11 W (including the back-end electronic board).

The WFM Camera coded mask is made of a 150 µm thick Tungsten foil and its pattern consists of 1040 × 16 elements of dimensions 250 µm × 14 mm with 2.4 mm spacing between the elements in the coarse resolution direction. The mask open fraction is 25%. The detector-mask distance is 202.9 mm to achieve the required angular resolution.

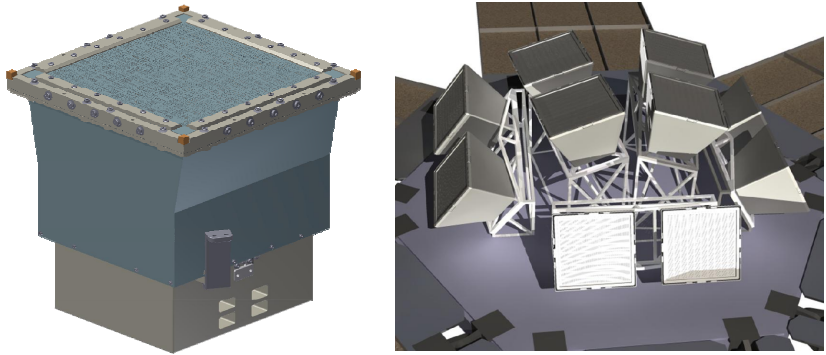

Figure 1 Mechanical design (left) of a WFM camera. The overall WFM design (right) is composed by 5 Units, each one made of 2 cameras with the fine resolution directions orthogonally arranged

The size of the mask is ~1.7 times larger than the detector, in order to achieve a flat (i.e., fully coded) region in the centre of the Field of View (FoV). In the LOFT WFM we can identify the following elements:

- the Camera: the assembly of a detector plane with its own collimator and mask, providing a fine (few arcmins) angular resolution in one coordinate and a coarse (few degrees) in the orthogonal direction;
- the Camera Unit: a set of 2 Cameras oriented at 90°, covering the same field of view. The use of the 2 Cameras as a Unit enables a fine 2D angular resolution;
- the WFM Assembly: the total set of Cameras composing the WFM, covering the entire field of view with fine angular resolution.

In Table 1 the Camera design is summarized.

Table 1 LOFT/WFM camera design

| Parameter | Value |
|---|---|
| Mask thickness | 150 µm |
| Mask material | Tungsten |
| Mask open fraction | 0.25 |
| Mask-detector distance | 202.9 |
| Mask physical size | 260 × 260 mm$^2$ |
| Mask pitch | 250 µm × 16.25 mm |
| On-axis effective area (@ 6 keV) | 41 cm$^2$ |

The whole WFM is composed of 5 Camera Units (10 Cameras, see Figure 1), each equipped with its Back-End Electronics, and a Data Handling Unit. The Back-End Electronics and Power Supply Unit (providing low, medium and high voltages power lines) for each Camera Unit are collected into a single electronics box, located close to the Camera Units. Each Camera is composed of a Detector Tray that includes 4 Silicon Drift Detectors with their 4 Front-End Electronics, a Be layer to shield the detectors from micrometeoroids, and a Coded Mask covered by a Thermal Blanket. The Units and the Cameras in the WFM are organized to achieve a high level of redundancy.

# 3. LOFT/WFM SIMULATOR

## 3.1 Simulation of astrophysical sources

In order to simulate an observation with the WFM, a sample of individual photons originating from the target sources is generated. For this purpose the tool 'phogen' from the SImulation of X-ray TElescopes[5] software package is used. It requires the specification of the observed targets in the SIMulationinPUT file format (SIMPUT).[6]

The SIMPUT file format is based on the FITS format[i] and provides the possibility to define either individual or a large number of sources. Each source is characterized by an energy spectrum and optional timing properties and spatial extent. The LOFT WFM SIMPUT files contain either an individual or multiple point-like X-ray sources with constant brightness and a Crab-like energy spectrum. The Cosmic X-ray Background (CXB) is taken into account by means of an additional SIMPUT file. The spectral model by Gruber et al. (1999)[7] is used.

Based on the source specifications and the pointing direction of the instrument, a sample of X-ray photons is produced for each simulation. The rate of generated photons is determined by the flux of the observed source and the geometric mask area of the WFM camera. Each photon is characterized by its energy, its arrival time at the instrument, and its direction of origin given in equatorial coordinates. In order to account for the decrease of projected area with respect to the geometric area of the mask for photons at non-zero off-axis angles, an algorithm is applied that discards individual generated photons list with a probability $p = 1 - \cos(\theta)$, where $\theta$ is the off-axis angle of the photon. This means that the rate of photons in the resulting data set is determined properly considering the projected mask area exposed to the X-ray source emission.

All generated photons are stored in a FITS table that can be processed through the instrument simulator.

## 3.2 LOFT/WFM detection plane simulations

The proposed LOFT/WFM detection plane is based on the Silicon Drift Detectors (SDDs). These detectors were initially developed for the Inner Tracking System (ITS) in the ALICE experiment of the Large Hadron Collider (LHC) at CERN,[8,9] by one of the scientific institutes involved in the LOFT Consortium – INFN Trieste, Italy – in co-operation with Canberra Inc. The key properties of the Silicon Drift Detectors[10] are their capability to read-out a large photon collecting area with a small set of low-capacitance anodes, allowing to reach excellent spectroscopic performance with very good imaging capabilities. The working principle of the SDD can be summarized as follows: the cloud of electrons generated by the interaction of an X-ray photon in the silicon bulk is focused on the middle plane of the detector, and then is drifted towards the read-out anodes, driven by a constant electric field sustained by a progressively decreasing negative voltage applied to a series of cathodes. During the drift time, the electron cloud size increases due to the diffusion mechanism.

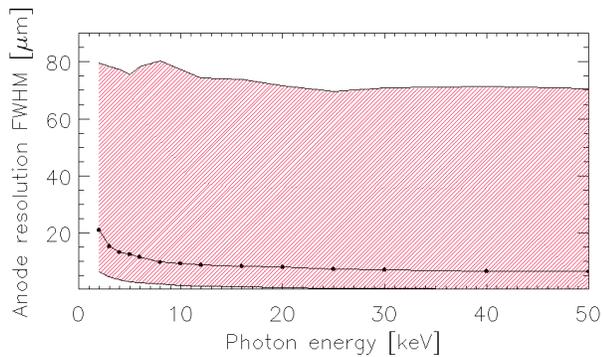 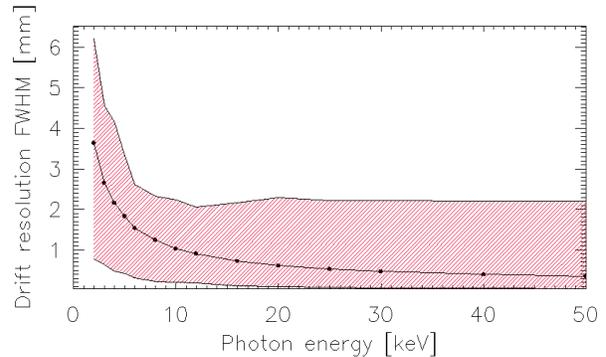

Figure 2. SDD anodic spatial resolution (FWHM) as a function of the photon energy for the LOFT/WFM detectors. The shaded area represents minimum-maximum spatial resolution performance inside the detector channel. Filled circles show the channel averaged values.

Figure 3. SDD drift spatial resolution (FWHM) as a function of the photon energy for the LOFT/WFM detectors. The shaded area represents minimum-maximum spatial resolution performance inside the detector channel. Filled circles show the channel averaged values.

---

[i] http://fits.gsfc.nasa.gov

We developed a Monte Carlo simulator describing the charge drift and diffusion inside the SDD to estimate the spectroscopic and imaging capabilities of the Silicon Drift Detectors. A detailed description of the linear SDD simulator can be found in Evangelista et al. (2012).[11] In Figure 2 and Figure 3 we show the simulated detector spatial resolution in the anodic and drift direction (hereafter fine and coarse directions, respectively) as a function of the incident photon energy.

The overall energy resolution of the SDD detector is affected by various sources of noise and systematics. The reconstruction of the energy of an incoming photon depends on the following factors:

- Fano noise (~118 eV FWHM at 6 keV)
- Electronic noise (ENC≤13 e⁻ rms)
- Common mode noise (25 e⁻ rms)
- Charge reconstruction (number of anodes that sample the charge cloud, common mode noise subtraction, reconstruction algorithm)
- ADC quantization noise.

The SDD simulator allowed us to simulate all the noise components listed above at different X-ray energies for whatever photon absorption position in the SDD sensitive area.

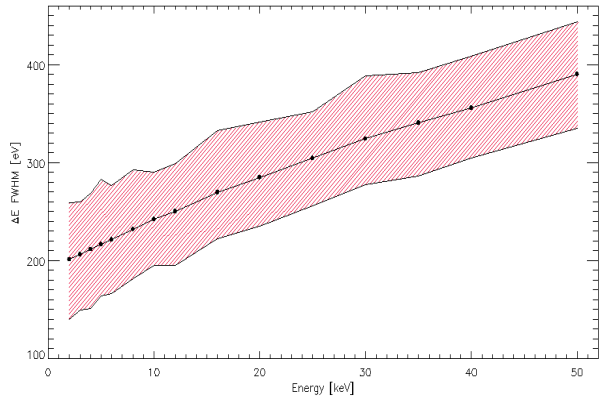

Figure 4. WFM energy resolution (FWHM) as a function of the photon energy. Solid circles represent the resolution values averaged on the whole SDD channel while the shaded area shows the range of energy resolution for each photon energy.

Figure 4 shows the simulated energy resolution (FWHM) as a function of the photon energy for the LOFT/WFM instrument. Solid circles represent the resolution values averaged on the whole SDD channel while the shaded area shows the range of energy resolution for each photon energy (the resolution depends on the photon absorption point).

### 3.3 Image reconstruction

In a coded mask instrument, the detector recorded picture array D can be expressed as

$$D = S \star M + N$$

where $\star$ is the correlation operator, $S$ is the array representing the sky image, $M$ is the mask and $N$ is some noise which is independent from the signal. We developed an image deconvolution procedure based on the widely used balanced cross-correlation (BC) method.[4] In the BC method, the deconvolved sky image can be obtained by correlating the detector image $D$ with a reconstruction array $R$, built in such a way that $M \star R$ approximate a δ function.

Using the reconstruction array, results that:

$$S = D \star R = S \star [M \star R] + N \star R.$$

The $R$ matrix is equal to 1 for each open mask element, and is equal to $t/(1-t)$ (where t=0.25 is the mask open fraction) for the closed (opaque) mask elements.[12,13]

The imaging procedure (written in IDL) takes as input:

- the photon list produced by the WFM simulator described in Section 3.1 (FITS binary table);
- the mask code (as a FITS image);
- a description of the WFM geometry including the reference system, the mask-to-detector distance, the mask thickness, etc. (FITS binary table);
- a description of the detector geometry and the location and dimensions of the SDD active areas (FITS image)

The sky images are obtained by cross-correlating the accumulated detector images with the reconstruction array described above. To take into account the non-sensitive detector regions (and the possible presence of switched-off channels in the detection plane) a "balancing" array $D_{bal}$ is used in the image cross-correlation. $D_{bal}$ is correlated with $R$ and then subtracted (after a proper normalization) to the sky image obtained with the cross correlation of $D \star R$. This allows us to subtract the low-frequency coding noise introduced by the detector non-sensitive regions.

The expected Point Spread Function of the system (SPSF) is triangle-shaped with a full width at half maximum equal to the width of one resolution element. It is worth noting that, as described in Fenimore and Cannon (1981),[14] a true $\delta$ function cannot be obtained in any case because an infinite frequency response would be required. Such a response is obviously not possible in a system with finite-sized holes.

Considering the detector resolution element size ($d$) and the fact that the SDD is a non-pixelated, continuous detector, a finely sampled cross-correlation procedure can be used to improve the image reconstruction.[14] In such a procedure the $D$ and $R$ arrays contain a number $r$ ($r>1$) of samples per resolution element. A value of $r = 16$ (which corresponds to a detector element size of ~16 μm × 1 mm) is used in the study of the SPSF in order to emphasize all the deviation from an ideal SPSF introduced, for example, by the inclined penetration and the mask vignetting.

At the end of the imaging procedure the software provides, as output, a number of FITS files including: the detector images, the deconvolved sky images, the variance images and the significance images

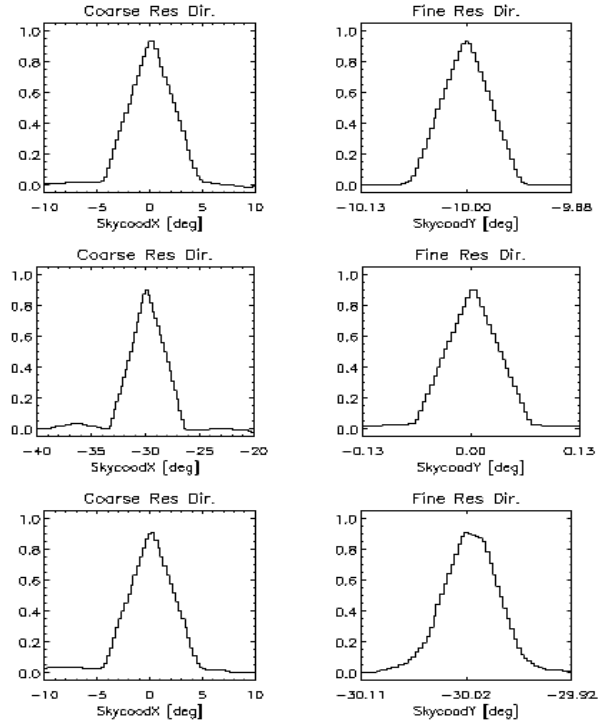

Figure 5. Finely sampled 2-20 keV SPSF along the coarse resolution direction (left panels) and along the fine resolution direction (right panels) for simulations of the Crab Nebula in different FoV positions.

Figure 5 shows the finely sampled (16 bins per resolution element) 2-20 keV SPSF along the coarse resolution direction (left panels) and along the fine resolution direction (right panels) for 5 simulations of the Crab Nebula located at off-axis angles ($\theta_x$, $\theta_y$) of (0°,0°), (-10°, 0°), (0°, -10°), (-30°,0°) and (0°, -30°). The exposure time in each simulation is 1 ks. In Figure 6 the effect of inclined penetration and vignetting is shown as a function of the photon energy for a Crab-like source located at $\theta_x = 0°$, $\theta_y = -30°$. Right panels show the SPSF in four energy bands (2-4 keV, 4-6 keV, 6-10 keV and 10-20 keV) for the fine resolution direction. The effect of vignetting is visible in the SPSF at low energies (first and second right panel) as a flattening of the PSF peak. The inclined penetration becomes more important as the photon energy increases. The penetration effect can be considered firstly as a "blurring" of the SPSF, due to the fact that the dependence of the mean photon absorption depth with the energy translates in an energy dependence of the actual mask-detector distance. Moreover, the peak position changes of ~1' between 2 and 20 keV, adding a large systematic error in the point source location accuracy for off-axis sources when the full energy band is considered in a simple deconvolution procedure. As a matter of fact, in a coded mask based instrument, this position shift can be effectively corrected by means, for example, of an Iterative Removal Of Sources (IROS) procedure.

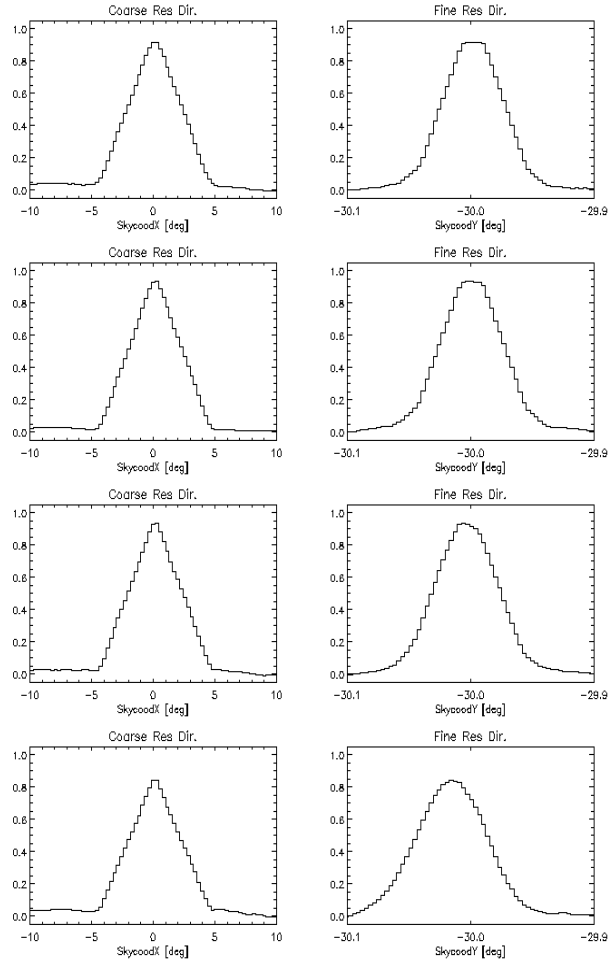

Figure 6. SPSF for a Crab like source 30° off-axis (fine resolution direction) in four energy bands (from top to bottom: 2-4 keV, 4-6 keV, 6-10 keV and 10-20 keV). The effects of the vignetting (flattening of the PSF) and of the inclined penetration (blurring of the PSF and drift of the source peak as a function of energy) are clearly visible.

### 3.4 Camera sensitivity

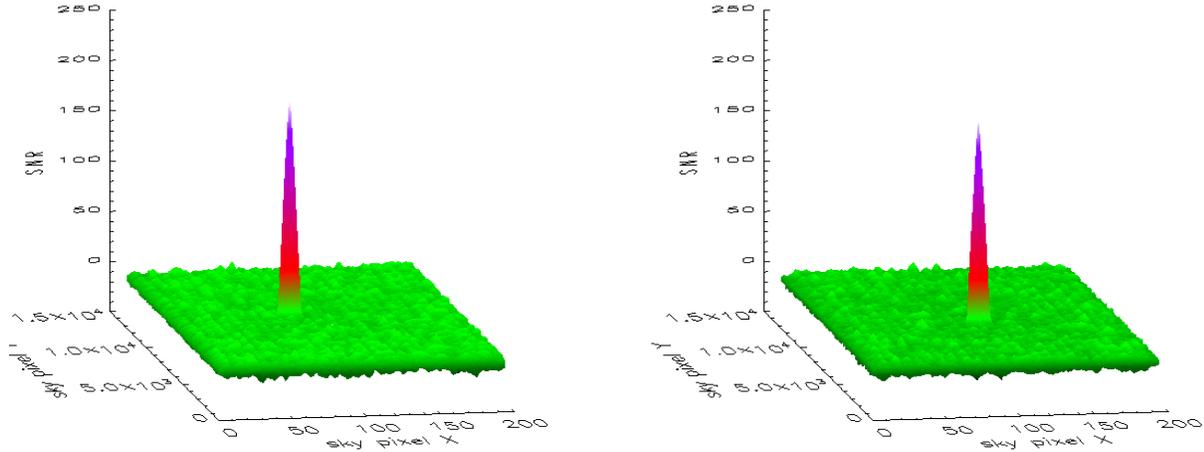

Figure 7 - Simulation of 1 ks long observation of the Crab Nebula located 10° off-axis along the coarse resolution direction (left panel, with $\theta_x = -10°$, $\theta_y = 0°$) and 10° off-axis along the fine resolution direction (right panel, with $\theta_x = 0°$, $\theta_y = -10°$). The images are represented in the detector reference system.

A number of simulations of an isolated Crab-like source, located in different places in the camera FoV, has been performed in order to estimate the point source sensitivity of the LOFT/WFM instrument.

All the simulations contain a 1 ks exposure of the Crab Nebula plus the Cosmic X-ray background. For the Crab Nebula, we considered an absorbed power-law spectrum with $\Gamma=2.1$, $nH=4\times10^{21}$ cm$^{-2}$ and a normalization of 9.5 ph cm$^{-2}$ s$^{-1}$ keV$^{-1}$ at 1 keV, while the background model is described in Section 3.1

Figure 7 shows the reconstructed significance images for the Crab Nebula located 10° off-axis in the coarse resolution direction (left panel: $\theta_x = 10°$, $\theta_y = 0°$) and 10° off-axis in the fine resolution direction (right panel: $\theta_x = 0°$, $\theta_y = -10°$) as represented in the detector reference system. The results of the analysis are summarized in Table 2 for the five simulations performed.

Table 2 Results of the simulations

| $\theta_x$ [deg] | $\theta_y$ [deg] | SNR (1ks) | 1 Camera Sens. (50 ks, 5σ) [mCrab] |
|---|---|---|---|
| 0 | 0 | 225.2 | 3.1 |
| -10 | 0 | 220.2 | 3.2 |
| 0 | -10 | 209.9 | 3.4 |
| -30 | 0 | 127.0 | 5.6 |
| 0 | -30 | 108.7 | 6.5 |

The single camera sensitivity is ~3 mCrab (50 ks, 5σ) for on-axis isolated sources. The sensitivity for off-axis sources is worsened by the mask vignetting, especially when the source lies off-axis in the fine-resolution direction ($\theta_y$). In Figure 8 we report the one camera sensitive area as a function of the off-axis angle (1 pixel = 1 deg) in the coarse (x) and fine (y) resolution directions. The effect of mask vignetting in the fine resolution direction is clearly visible.

### 3.5 Imaging simulations for a camera unit and overall WFM performance

The imaging properties of two orthogonal Cameras (one Unit) have been studied in detail by means of the Monte Carlo simulator. A simulated 10 ks observation of the Galactic Centre is shown in Figure 9. The map represents a central region (about 40° × 30°) of the FoV of one Unit obtained by combining the sky simultaneously observed by the two Cameras.

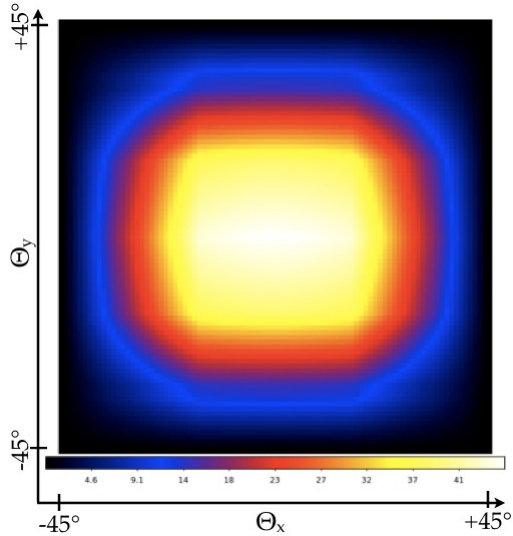 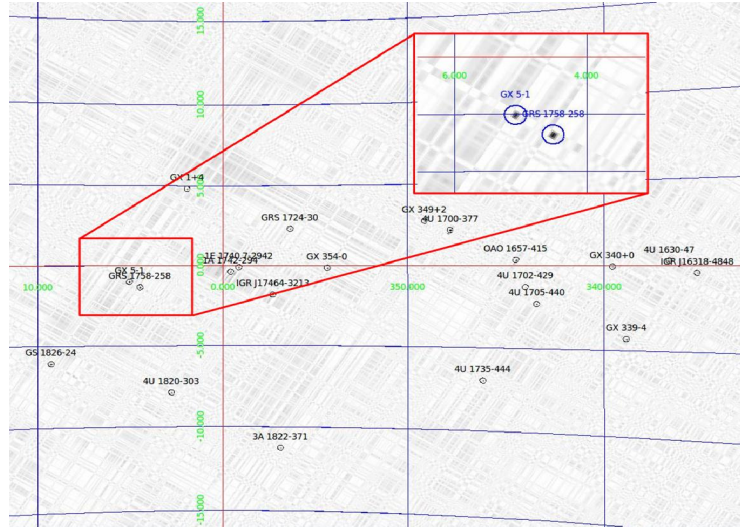

Figure 8 Map of the single camera sensitive area expressed in cm$^2$. The map takes into account the main geometrical effects (mask open fraction, vignetting, shadowing of the collimator walls, detector non-sensitive areas, cos(θ) effect).

Figure 9 Unit simulation of a 10 ks observation of the Galactic Center. The map shows a central region (about 40° × 30° large) of the Unit FoV.

The overall performance of the LOFT Wide Field Monitor are summarized in Table 3.

Table 3 Overall WFM performance

| Parameter | Camera | Unit | Overall WFM |
|---|---|---|---|
| Energy range | 2-50 keV | 2-50 keV | 2-50 keV |
| Active detector area | 182 cm$^2$ | 364 cm$^2$ | 1820 cm$^2$ |
| Peak effective area | 45 cm$^2$ | 90 cm$^2$ | 90 cm$^2$ |
| FoV Zero response | 90° × 90° | 90° × 90° | 210° × 90° + 90° × 90° |
| FoV 20% response | 60° × 60° | 60° × 60° | 180° × 60° + 60° × 60° |
| Angular resolution | 5' × 5° | 5' × 5' | 5' × 5' |
| Point Source Location Accuracy (10σ) | <1' × 30' | <1' × 1' | <1' × 1' |
| On-axis sens. 5σ, 3 s, Gal. center | 380 mCrab | 270 mCrab | 270 mCrab |
| On-axis sens. 5σ, 10 ks, Gal. center | 6.6 mCrab | 4.7 mCrab | 4.7 mCrab |

## 4. CONCLUSIONS

We developed a comprehensive mixed analytic/Monte Carlo simulator in order to optimize and characterize the response of the coded-mask Wide Field Monitor on-board the candidate ESA M3 mission LOFT.

By means of a detailed physical description of the WFM elements (coded mask, SDD detection plane and front-end electronics), the software allows us to simulate the instrument response to both astrophysical point sources and diffuse X-ray background, for various exposure times and several different source positions in the FoV. The results obtained with the simulator show that a lightweight, compact, low-power and sensitive imager can be effectively built exploiting the coded-aperture imaging principle coupled with a detection plane based on the large area Silicon Drift Detectors technology. The instrument will cover a large FoV in the same energy range as the Large Area Detector, namely almost 50% of the sky in 2-50 keV energy range. Its design would allow the detection of transient and steady sources with fluxes down to a few mCrab per day, with a spectral resolution better than 300 eV FWHM at 6 keV, a timing resolution of about 10 μs and an angular resolution of 4.7', which translates into a sub-arcmin point-source location accuracy (PSLA) for high significance (>10σ) detections.